\documentclass[
prd
,showpacs,amssymb,superscriptaddress,aps
]{revtex4-2}
\usepackage{graphicx}
\usepackage{color}
\input{epsf}

\usepackage{amsmath,amssymb}
\usepackage{bm}
\usepackage{times}
\usepackage{ulem}

\newcommand{\dalm}{\kern1pt\vbox{\hrule height 0.9pt\hbox{\vrule width 0.9pt
\hskip 2.5pt\vbox{\vskip 5.5pt}\hskip 3pt\vrule width 0.3pt}\hrule height 0.3pt}
\kern1pt}

\newcommand{\lsim}{\, \, \raisebox{-0.8ex}{$\stackrel{\textstyle <}{\sim}$ }}

\begin{document}



\title{Universality in quasinormal modes of neutron stars with quark-hadron crossover}

\author{Hajime Sotani}
\email{sotani@yukawa.kyoto-u.ac.jp}
\affiliation{Astrophysical Big Bang Laboratory, RIKEN, Saitama 351-0198, Japan}
\affiliation{Interdisciplinary Theoretical \& Mathematical Science Program (iTHEMS), RIKEN, Saitama 351-0198, Japan}

\author{Toru Kojo}
\affiliation{Department of Physics, Tohoku University, Sendai 980-8578, Japan}


\date{\today}

\begin{abstract}
We examine the gravitational wave frequencies of the fundamental ($f$-) and 1st pressure ($p_1$-) modes excited in the neutron star models constructed with the quark-hadron crossover (QHC) type equations of state (EOS). We find that the $f$-mode frequencies with QHC EOS basically are smaller and the $p_1$-mode frequencies with QHC EOS are larger than those with hadronic EOS, focusing on the neutron star model with a fixed mass. We also find that the universality in the $f$-mode frequencies multiplied by the stellar mass as a function of the stellar compactness or as a function of the dimensionless tidal deformability, which is derived with various hadronic EOSs, can keep even with QHC EOS. 
That is, using these universal relations, one cannot distinguish QHC EOS from hadronic EOSs.
Instead, using the relations one can extract the stellar radii whose evolution from low to high mass neutron stars can differentiate QHC from hadronic EOSs. 
On the other hand, we find that the $p_1$-mode frequencies multiplied by the stellar mass with QHC EOS significantly deviate in a certain mass range from the corresponding empirical relations derived with various hadronic EOSs, with which one may distinguish QHC EOS from hadronic EOSs. 

\end{abstract}

%
\maketitle


\section{Introduction}
\label{sec:I}

Neutron star, which is a massive remnant left after a supernova explosion, realizes extreme environment \cite{ST83}. 
The density inside a 
star significantly exceeds the nuclear saturation density, $n_0 \simeq 0.16\, {\rm fm}^{-3}$, while the magnetic and gravitational fields become much stronger than those observed in our solar system. Thus, one can probe physics under such extreme states by observing the neutron stars and/or their phenomena. In fact, after the discovery of the massive neutron stars, whose masses are around $2M_\odot$, soft equations of state (EOSs) have been ruled out \cite{D10,A13,C20,F21}. The observation of the gravitational wave produced by colliding neutron stars, GW170817 \cite{gw170817}, tells us the tidal deformability, which leads to the constraint on the $1.4M_\odot$ neutron star radius \cite{Annala18,Capano20,Dietrich20}. One of the relativistic effects, i.e., the light bending, is also an important phenomenon to extract the neutron star properties. Owing to such an effect in the strong gravitational field induced by the neutron star itself, the photon radiating from the neutron star's surface can bend, which leads to the modulation of the pulsar light curve. So, by carefully observing the pulsar light curve, one could primarily constrain the neutron star compactness, which is the ratio of the stellar mass to its radius (e.g., \cite{PFC83,LL95,PG03,PO14,SM18,Sotani20a}). In practice, the neutron star mass and radius, i.e., PSR J0030+0451 \cite{Riley19, Miller19} and PSR J0740+6620  \cite{Riley21, Miller21}, have been constrained through observation by the Neutron Star Interior Composition Explorer (NICER) equipped on the International Space Station. On the other hand, the nuclear properties in a relatively lower density region, e.g., less than at most twice the saturation density, have been gradually constrained via terrestrial experiments, which also leads to the constraints on the neutron star mass and radius \cite{SNN22,SO22,SN23}.

To extract the neutron star properties, the oscillation frequency (or gravitational waves) from a neutron star is another important information. In general, an object has its own specific frequencies depending on its interior conditions. So, one may see the interior properties as an inverse problem, if one would detect such specific frequencies. This technique has already been established as seismology on Earth and helioseismology on Sun. In the same way, one may extract the properties of a compact object by observing its oscillation signal, which is known as asteroseismology (or gravitational wave asteroseismology if using gravitational waves). In practice, by identifying the quasi-periodic oscillations observed in the magnetar flares with the neutron star crustal oscillations, one could constrain the crust properties and/or the stellar model (e.g., \cite{SNIO2012,SIO2016,SKS2023}). In a similar way, once one would detect the gravitational waves from a (isolated) compact object, one could see the stellar mass, radius, and EOS (e.g., \cite{AK1996,AK1998,STM2001,SH2003,SYMT2011,PA2012,Sotani2020,Sotani21,SD2021}).

The neutron star mass and radius have been becoming constrained through astronomical observations and terrestrial experiments, but the state of matter is still quite uncertain in a higher-density region. 
In particular, 
the existence of quark matter inside a neutron star is still under intensive debate.
But quark degrees of freedom should be important in high-mass neutron stars. The inferred core baryon density reaches $\sim 3$-$5n_0$ where baryons may overlap.
At least two possibilities are available for hadron-to-quark matter transitions.
One is a first-order phase transition from hadronic matter to quark matter, where Gibbs phase equilibrium conditions are imposed. 
This approach is more historical and there are many studies. In the region between a pure hadronic phase and a pure quark phase, 
the so-called mixed phases must exist \cite{HPS93,Maruyama07}, which is a similar structure to the crustal pasta structure. 
We call these models collectively 1st PT models.
However, in general, this type of EOS often tends to be too soft to support massive neutron stars. Another possibility is the quark-hadron crossover (QHC) type of EOS, which is analogous to the BCS-BEC crossover in a many-body system of ultra-cold atoms. 
The crossover picture has been motivated by observing that 
hadronic matter in which baryons interact by meson or quark exchanges, and strongly correlated quark matter which is about to get confined, are difficult to distinguish, where the effective degrees of freedom continuously change.
The microscopic models have been recently constructed in Refs. \cite{kojo21,FKM23}. A more phenomenological but practical construction of crossover EOS is to interpolate hadronic and quark EOSs in a way consistent with the thermodynamic stability and causality conditions \cite{Masuda13,QHCreview}. The advantage of the crossover picture over 1st PT models is that, unlike in 1st PT models, we do not have to demand quark EOS to be softer than {\it extrapolated} hadronic EOS, which is untrustable at high density; as a result stiff quark EOS can be a candidate of EOS. Stiff quark matter is the key to explaining the two-solar mass constraints while keeping microscopic descriptions natural.

An important question is how to observationally discriminate the crossover scenario from the first-order transition and purely hadronic scenarios. One of the outstanding features in the crossover model is the rapid stiffening of EOS which begins to occur at $\sim 1.5$-$3 n_0$ and subsequent relaxing to the conformal behaviors at higher density. This evolution of stiffness is reflected in a peak of the sound velocity. Such a peak is absent in the other scenarios: purely hadronic EOS shows gentle and monotonic evolution of stiffness while 1st PT models lead to vanishing sound velocity for some density interval.
The efficient way to see these differences is to study the evolution of neutron star radii for the mass range $\gtrsim 1.4M_\odot$. Early stiffening in the crossover scenario leads to the radii being larger than in the other scenarios.
In this context, the $f_2$-frequency of gravitational waves from binary neutron star mergers, which are sensitive to the radii of the high mass region,
has been studied in numerical relativity \cite{Yongjia22}. 
Here, we note that $f_2$-frequency is one of the distinct peak frequencies appeared in the power spectrum from the rotating hypermassive neutron stars produced after a merger of binary neutron stars \cite{BJ12,TRB14}, which is different from the fundamental ($f$-) modes in gravitational wave examined in this study. 
It was argued that the $f_2$ in different scenarios can be distinguished in the third generation detectors planned after 2035 \cite{ET,CE}.

Concerning the mass and radius of a neutron star, several universal relations among observables have been proposed. 
For purely hadronic EOS, the universal relations for the combinations of observables hold independently from EOSs.
For example, the I-Love-Q relation, which is the relation between the moment of inertia, the Love number, and the quadrupole moment independently of the interior structure of neutron stars (and quark stars), is one of the universal relations \cite{Yagi13}. In a similar fashion, several universal relations in quasinormal modes of neutron stars have been found (e.g., \cite{AK1996,AK1998,TL05,Sotani2020,Sotani21,ZL22}). 
But how universal those relations are needs clarification, as they are mostly derived from the family of hadronic EOSs. 
At least, the 1st PT models seem to be consistent with the universal relation \cite{MLZH20,ZL22}.
Thus, it is natural to ask whether crossover models follow the same universality as purely hadronic models.
The universal relation can be used in two-fold ways.
If the relation is violated in crossover models, one can characterize crossover EOS with the violation.
Instead, if holds, one can use the relation to extract, e.g., the radius of a neutron star, and study the trend differing from hadronic baselines.
In this study we examine the gravitational wave frequencies from the neutron stars constructed with QHC EOSs \cite{QHC19,QHC21} and examine the universality. 
We note that the amplitude of the gravitational wave cannot be directly derived from our study, based on a linear perturbation analysis. Even so, if the energy of the gravitational wave with a focusing oscillation mode is more than $10^{-10}M_\odot$, one may detect the gravitational waves from even an isolated neutron star located in our galaxy.

This manuscript is organized as follows. In Sec. \ref{sec:EOS}, we mention the neutron star models considered in this study together with the adopted EOSs. In Sec. \ref{sec:perturbations}, we show the gravitational wave frequencies from neutron stars constructed with QHC EOSs and discuss their universality. Finally, in Sec. \ref{sec:Conclusion}, we conclude this study. Unless otherwise mentioned, we adopt geometric units in the following, $c=G=1$, where $c$ and $G$ denote the speed of light and the gravitational constant, respectively.

\section{EOS and Equilibrium models}
\label{sec:EOS}

In this study, we simply consider a non-rotating and spherically symmetric neutron star as an equilibrium model. The metric describing such an object is given by
\begin{equation}
  ds^2 = -e^{2\Phi}dt^2 + e^{2\lambda}dr^2 + r^2\left(d\theta^2 + \sin^2\theta d\phi^2\right), \label{eq:metric}
\end{equation}
where the metric functions, $\Phi$ and $\lambda$, depend on only the radial coordinate, $r$. The enclosed gravitational mass, $m$, inside the radial position, $r$, is directly associated with $\lambda$ through $2\lambda=-\ln(1-2m/r)$. Assuming that a neutron star is composed of a perfect fluid, one can construct the stellar models by integrating the Tolman-Oppenheimer-Volkoff equation together with an appropriate EOS for neutron star matter.

In this study, we focus on only the family of QHC21T EOSs for studying the gravitational wave frequencies from neutron stars (see Ref.~\cite{QHC21} for details about QHC21T EOSs). We note that there are several other QHC EOSs, such as the family of QHC19 \cite{QHC19} and the EOS model proposed by Li, Yan \& Ping \cite{LYP22}, except for QHC21. QHC21T is constructed as follows. Togashi EOS \cite{Togashi17}, which is a pure hadronic EOS, is adopted in a lower-density region, i.e., $n_{\rm b}\le 1.5n_0$, where $n_{\rm b}$ and $n_0$ denote the baryon number density and it at the saturation point for the symmetric nuclear matter ($n_0=0.160$ fm$^{-3}$ for Togashi EOS corresponds to the saturation density $\rho_0=2.68\times 10^{14}$ g/cm$^3$). Togashi EOS is derived by a variational method, employing the Argonne v18 (AV18) two-body nuclear potential and the Urbana IX (UIX) potential for the three-body nuclear force. Meanwhile, the quark matter EOS calculated with the Nambu-Jona-Lasinio (NJL) model is adopted in a higher-density region, i.e., $n_{\rm b}\ge 3.5n_0$. The quark matter EOS has two parameters characterizing the stiffness of quark matter, i.e., the coupling constant $g_{\rm V}$, which is associated with the short-range density-density repulsion strength in quantum chromodynamics (QCD), and coupling constant $H$, which describes an attraction between quarks. And then, the region with $1.5n_0\le n_{\rm b}\le 3.5n_0$ corresponds to the quark-hadron crossover region. In this region, the pressure, $p$, is simply assumed as a fifth-order function of baryon chemical potential, $\mu_{\rm B}$, where the coefficients in the fifth-order function are determined in such a way that $p$, $n_{\rm b}=\partial p/\partial \mu_{\rm B}$, and the susceptibility $\chi_{\rm B}=\partial^2 p/\partial \mu_{\rm B}^2$ should be continuous at $n_{\rm b}=1.5n_0$ and $3.5n_0$ \cite{note}. In addition to QHC21T EOSs, for reference, we also consider the neutron star models constructed with Togashi EOS.

\begin{figure}[tbp]
\begin{center}
\includegraphics[scale=0.6]{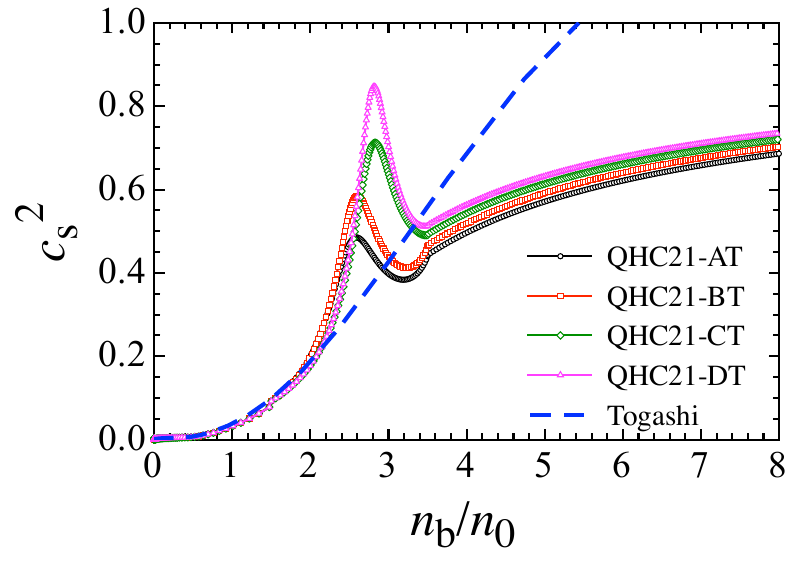}
\end{center}
\caption{
Profile of square of the sound velocity for the EOSs adopted in this study, where $n_{\rm b}$ and $n_0$ in the horizontal axis denote the baryon number density and $n_{\rm b}$ at the saturation point, respectively.
}
\label{fig:cs2}
\end{figure}

In particular, we adopt four specific QHC21T EOSs, i.e., 
QHC21-AT, -BT, -CT, and -DT whose coupling parameters, $(g_{\rm V},H)$, are shown in Table 1 in Ref. \cite{QHC21}. 
The adopted quark matter EOS becomes stiffer in the order of 
AT, BT, CT, DT, i.e., the quark matter EOS for QHC21-AT is the softest, while that for QHC21-DT is the stiffest among these four QHC21T EOSs. The stiffness of quark matter affects the peak of the sound velocity inside the crossover region. In fact, the maximum sound velocity becomes smaller in the order of 
AT, BT, CT, DT, as shown in Fig.~\ref{fig:cs2}, where we also show the sound velocity of the Togashi EOS for reference. We note that the presence of the peak (or local maximum) of the sound velocity at finite baryon density is one of the important features in QHC EOS (e.g., \cite{Masuda13,KS22}).

\begin{figure}[tbp]
\begin{center}
\includegraphics[scale=0.6]{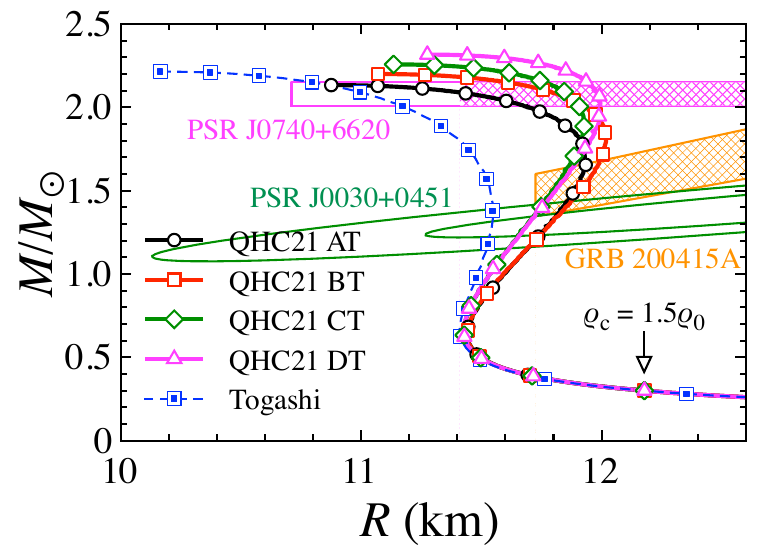}
\end{center}
\caption{
The mass and radius relation for the neutron star models constructed with QHC21T and with Togashi EOSs. In this figure, the leftmost stellar model for each sequence corresponds to the neutron star model with the maximum mass, while the stellar model indicated by an arrow is the neutron star model whose central density is $1.5\rho_0$. In addition, as a reference, we also plot some constraints obtained from the astronomical observations (see the text for details). 
}
\label{fig:MR}
\end{figure}

In Fig.~\ref{fig:MR}, we show the mass and radius relation for the neutron star models constructed with QHC21T and with Togashi EOSs. 
The marks in the curves correspond to specific core densities; for each stellar model, we set the step size by equally dividing the logarithmic density from $\rho=1.5\rho_0$ for QHC21T (and $\rho=\rho_0$ for Togashi) up to that at the maximum mass. Some of the stellar models constructed with lower central density using Togashi EOS are out of the figure. The marks shown in the following Figs.~\ref{fig:MR-Lam} -\ref{fig:Lam-fp1} are set in the same way.
In Fig.~\ref{fig:MR}, we also show some constraints obtained from astronomical observations. That is, the mass and radius constraints are shown for PSR J0030+0451 \cite{Riley19, Miller19} and PSR J0740+6620  \cite{Riley21, Miller21} with NICER observations, and for GRB 200415A by identifying the observed quasi-periodic oscillations with crustal torsional oscillations consistently with the terrestrial experimental data \cite{SKS2023}. We note that the neutron star models shown in Fig.~\ref{fig:MR} (or the EOSs adopted in this study) satisfy the constraint on the $1.4M_\odot$ neutron star radius obtained from the observations of GW170817, i.e., $R_{1.4} \le 13.6$ km \cite{Annala18}.

In addition to the stellar mass and radius, the dimensionless tidal deformability, $\Lambda$, due to the gravitational field induced by the companion star may be an important property in a binary system, because $\Lambda$ could be estimated from the chirp signal of gravitational waves just before the coalescence of a binary neutron star. $\Lambda$ is associated with the dimensionless quadrupole tidal Love number, $k_2$, through
\begin{equation}
  \Lambda = \frac{2}{3}k_2\left(\frac{M}{R}\right)^{-5},
\end{equation}
while $k_2$ is calculated from the integration of the ordinary differential equation shown in Ref. \cite{Hinderer08}. It has been shown that $\Lambda$ can be expressed as a function of the stellar compactness, $M/R$, almost independently of the EOSs (including Togashi EOS) \cite{Yagi14,Sotani21}. In Fig.~\ref{fig:MR-Lam}, in a similar way, we plot $\Lambda$ as a function of $M/R$ for the stellar models constructed with QHC21T and Togashi EOSs, from which one can observe that the universality between $\Lambda$ and $M/R$ can keep even with the QHC EOS.

\begin{figure}[tbp]
\begin{center}
\includegraphics[scale=0.6]{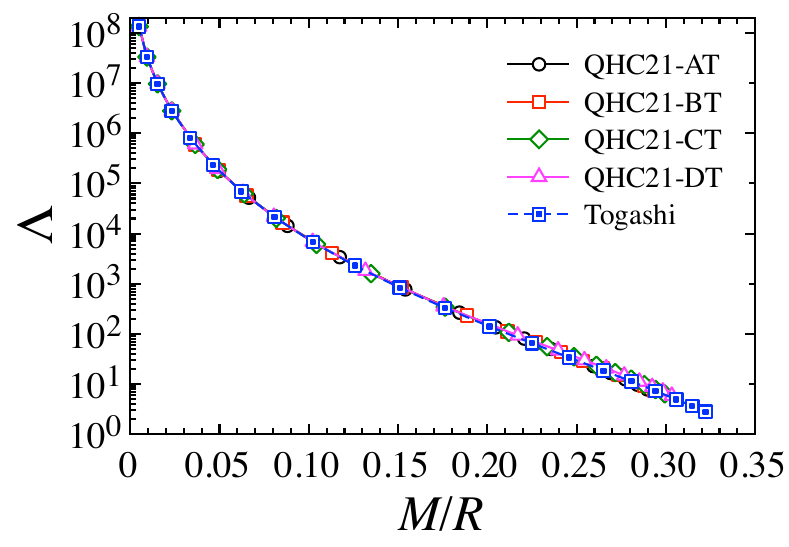}
\end{center}
\caption{
The dimensionless tidal deformability, $\Lambda$, is shown as a function of the stellar compactness, $M/R$, for the stellar models constructed with QHC21T and Togashi EOSs. We note that $M/R=0.17$ for the canonical neutron star with $1.4M_\odot$ and 12 km. 
}
\label{fig:MR-Lam}
\end{figure}

\section{Eigenfrequencies and universality}
\label{sec:perturbations}

To determine the eigenfrequencies of gravitational waves radiated from the neutron stars, one has to solve the eigenvalue problem. One can derive the perturbation equations by adding the metric and fluid perturbations to the equilibrium model and linearizing the Einstein equation. Then, one has to impose the appropriate boundary conditions, i.e., the regularity condition at the center, the vanishment of the Lagrangian perturbation of pressure at the stellar surface, and the condition of the purely outgoing waves at spatial infinity. The perturbation equations and how to deal with the boundary condition at spatial infinity are concretely shown in Refs. \cite{STM2001,ST20}. Since gravitational waves carry out the oscillation energy from the objects, the resultant eigenfrequencies become complex values, i.e., quasinormal modes, where the real and imaginary parts respectively correspond to the oscillation frequency and damping rate. But, the damping rate for the gravitational waves induced by the fluid motion is generally much smaller than the oscillation frequency. So, in this study, we adopt the zero-damping approximation, i.e., the imaginary part of the eigenfrequencies is set to zero \cite{Sotani22}.

\begin{figure}[tbp]
\begin{center}
\includegraphics[scale=0.6]{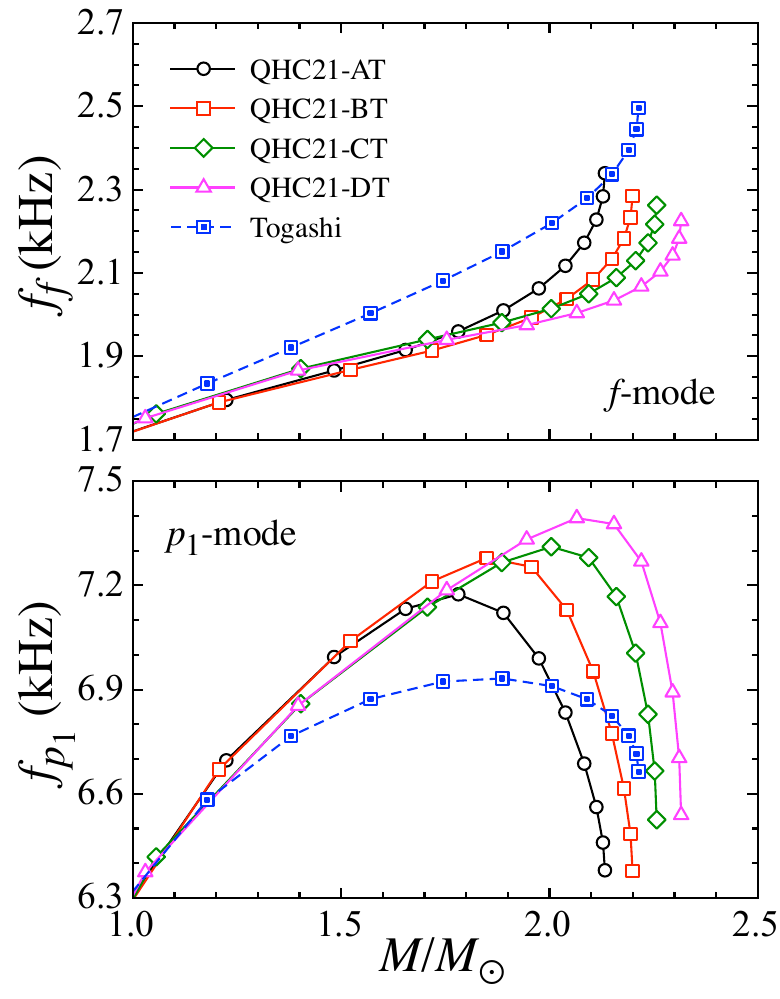}
\end{center}
\caption{
The $f$- and $p_1$-mode frequencies are shown as a function of the stellar mass in the top and bottom panels, respectively. 
}
\label{fig:fp-M1}
\end{figure}

In Fig.~\ref{fig:fp-M1}, the resultant frequencies of the $f$- and the 1st pressure ($p_1$-) modes are shown as a function of the mass of the neutron stars constructed with QHC21T and Togashi EOSs. From this figure, one can observe that the $f$-mode frequencies expected with QHC21T EOSs basically become smaller than those with Togashi EOS. On the other hand, the $p_1$-mode frequencies with QHC21T can become larger than those with Togashi within a certain mass range, depending on the stiffness of the EOS for quark matter. We also find that the stellar models with QHC21T EOSs, whose $p_1$-mode frequencies become maximum, roughly correspond to those with the local maximum radius, whose mass is $\sim 1.7-2.1M_\odot$. In addition, the maximum frequency of the $p_1$-mode becomes higher, as the peak sound velocity in QHC21T becomes larger (or as the stiffness of quark matter EOS increases).

\begin{figure*}[tbp]
\begin{center}
\includegraphics[scale=0.6]{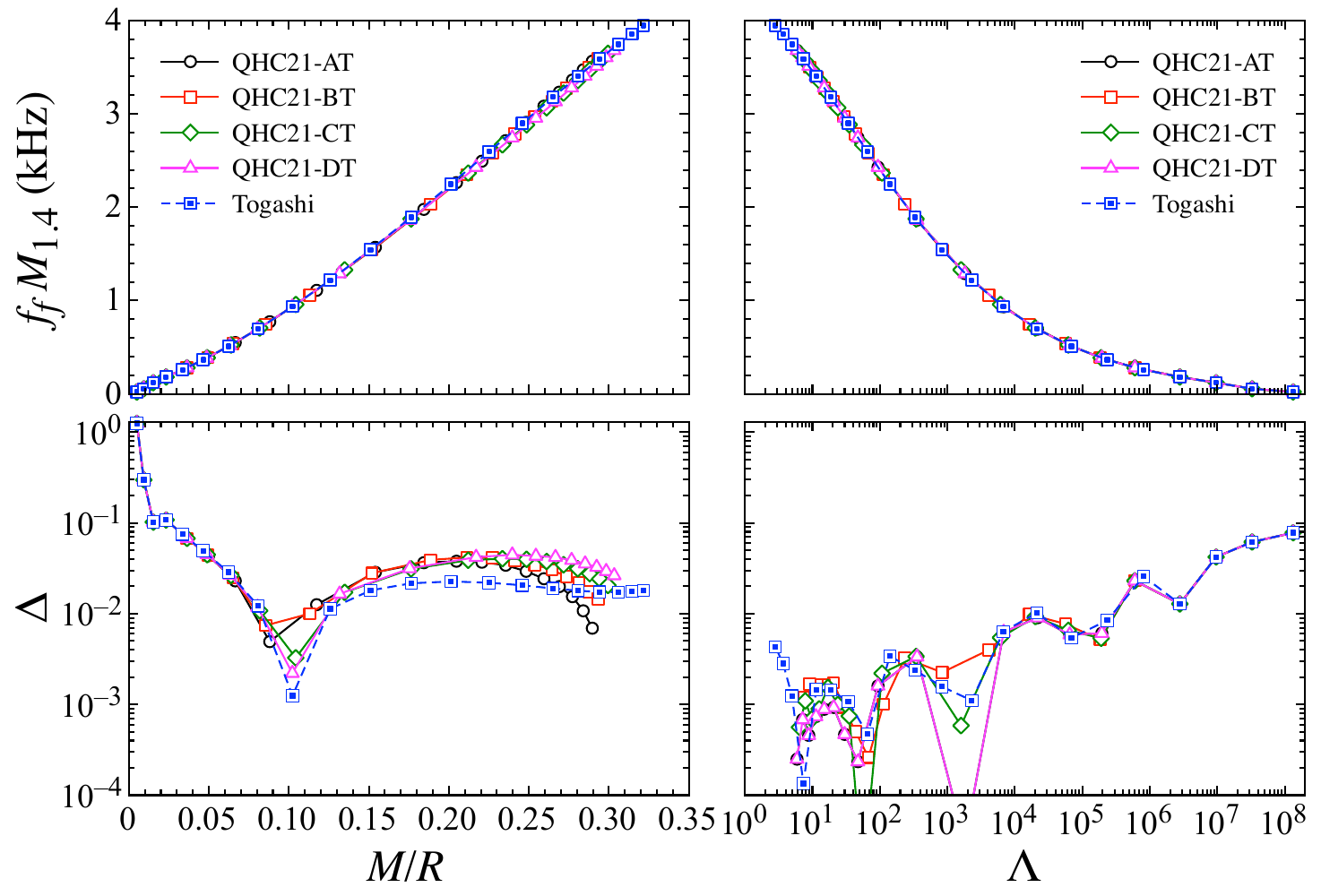}
\end{center}
\caption{
The $f$-mode frequencies multiplied by $M_{1.4}\equiv M/1.4M_\odot$ are shown as a function of $M/R$ (top-left panel) and $\Lambda$ (top-right panel) for the neutron star models constructed with QHC21T and Togashi EOSs. The bottom panels are the relative deviation in the value of $f_fM_{1.4}$ from the fitting formula, given by Eq. (12) in Ref. \cite{Sotani2020} and Eq. (7) in Ref. \cite{Sotani21}.
}
\label{fig:MR-Lam-ff}
\end{figure*}

Next, we focus on the universal relations with the quasinormal modes of neutron stars. Using the $f$-mode gravitational waves from the neutron stars constructed with various hadronic EOSs, it is derived that the universal relation of the $f$-mode frequencies, $f_f$, multiplied by the stellar mass, $M$, can be expressed as stellar compactness, $M/R$, given by Eq. (12) in Ref. \cite{Sotani2020} or as a function of the dimensionless tidal deformability, $\Lambda$, given by Eq. (7) in Ref. \cite{Sotani21}, i.e., the $f_fM-M/R$ and $f_fM-\Lambda$ relations. In Fig.~\ref{fig:MR-Lam-ff}, we similarly plot $f_fM_{1.4}$ as a function of $M/R$ in the top-left panel and as a function of $\Lambda$ in the top-right panel, where $M_{1.4}$ is the normalized stellar mass defined as $M_{1.4}\equiv M/1.4M_\odot$. In bottom panels, we show the relative deviation, $\Delta$, from the fitting formulae calculated as $\Delta = |{\cal A}_{\rm N} - {\cal A}_{\rm F}|/{\cal A}_{\rm N}$, where ${\cal A}_{\rm N}$ and ${\cal A}_{\rm F}$ respectively denote the value of $f_{f}M_{1.4}$ determined numerically from the eigenvalue problem and that estimated with the fitting formulae.
From this figure, one can observe that the universality of the $f_fM-M/R$ and $f_fM-\Lambda$ relations obviously keeps even with QHC21T EOSs, even though the deviation with QHC21 from the $f_fM_{1.4}-M/R$ relation is a little larger than that with Togashi. Namely, as long as the sound velocity is continuous inside the neutron star, the profile of sound velocity may be irrelevant in the universality of the $f_fM-M/R$ and $f_fM-\Lambda$ relations.

\begin{figure}[tbp]
\begin{center}
\includegraphics[scale=0.6]{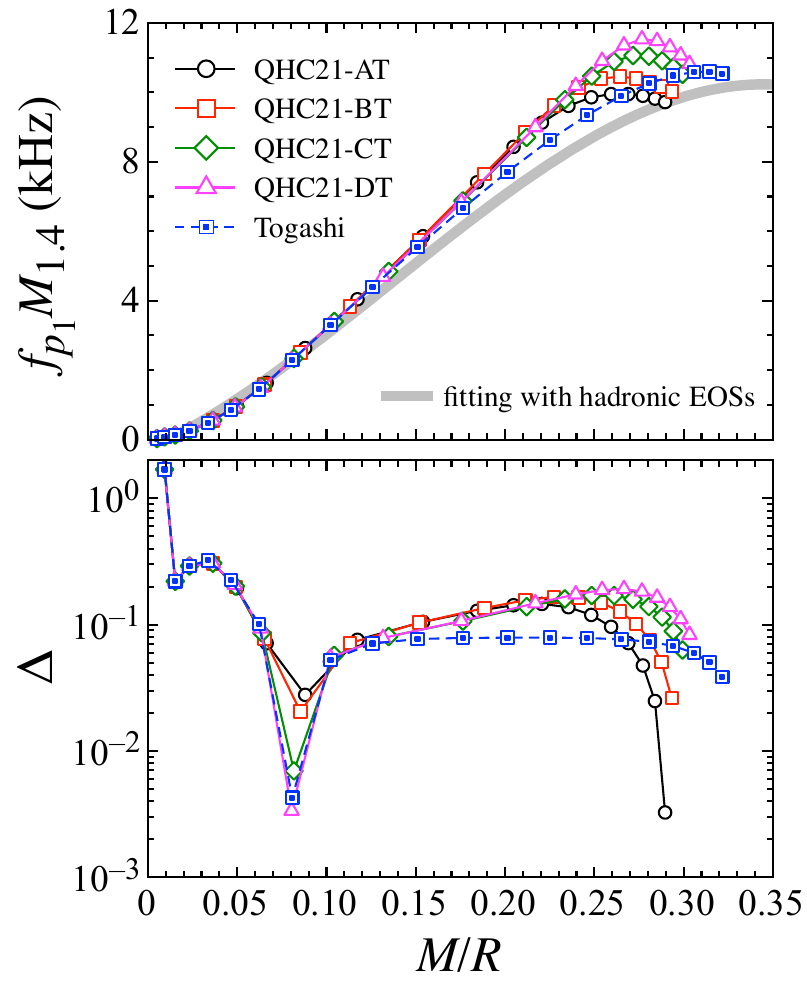}
\end{center}
\caption{
The $p_1$-mode frequencies multiplied by the normalized stellar mass, $f_{p_1}M_{1.4}$, are shown as a function of $M/R$ in the top panel, where the thick solid line denotes the fitting formula derived with various hadronic EOSs, given by Eq. (16) in Ref. \cite{Sotani2020}. We plot the relative deviation from the fitting formula in the bottom panel. 
}
\label{fig:MR-fp1}
\end{figure}

\begin{figure}[tbp]
\begin{center}
\includegraphics[scale=0.6]{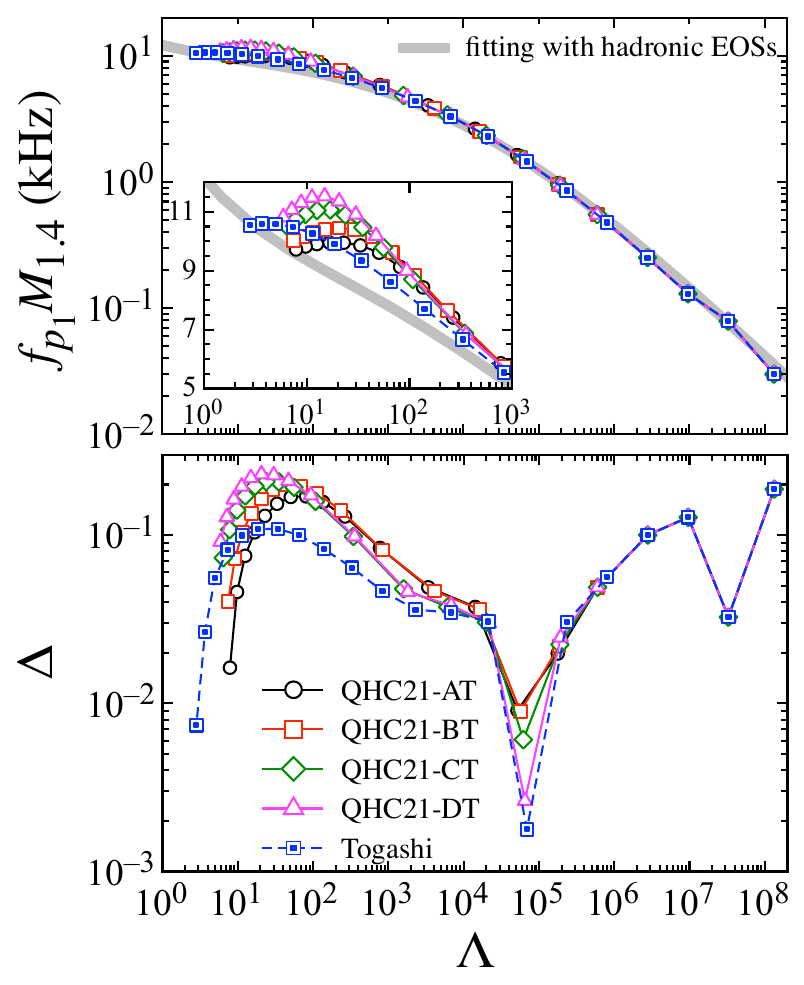}
\end{center}
\caption{
Same as in Fig.~\ref{fig:MR-fp1}, but the fitting formula plotted in the top panel is given by Eqs. (11) and (12) in Ref. \cite{Sotani21}. In the top panel, we also show the enlarged view focusing on the canonical neutron star models, whose value of $\Lambda$ is less than 1000. We note that the vertical axis in the enlarged view is a linear scale, instead of a log scale.
}
\label{fig:Lam-fp1}
\end{figure}

Finally, we discuss the $p_1$-mode frequencies, $f_{p_1}$, in gravitational waves. Compared to the $f$-mode frequencies, the $p_1$-mode frequencies cannot be expressed well independently of the EOSs. Nevertheless, the fitting formulae, which weakly depend on the EOSs, have been derived with the $p_1$-mode frequencies from the neutron star models constructed with various hadronic EOSs. In fact, in a similar way to the $f$-mode frequencies, $f_{p_1}M_{1.4}\ {\rm (kHz)}$ can be expressed as a function of stellar compactness as Eq. (16) in Ref. \cite{Sotani2020}
or as a function of the dimensionless tidal deformability as Eqs. (11) and (12) in Ref. \cite{Sotani21}.
For the case of the hadronic EOSs, one can estimate the value of $f_{p_1}M_{1.4}\ {\rm (kHz)}$ within less than $10\%$ accuracy independently of EOSs, using these fitting formulae \cite{Sotani2020,Sotani21}. In order to see how well these fitting formulae for $f_{p_1}$ work, we plot similar relations in Figs.~\ref{fig:MR-fp1} and \ref{fig:Lam-fp1}, where the thick solid line in the top panel denotes the expected value using the fitting formulae derived from hadronic EOSs. In the top panel of Fig.~\ref{fig:Lam-fp1}, we also show the enlarged view focusing on the canonical neutron star models, whose value of $\Lambda\lsim 1000$. We note that the vertical axis in this enlarged view is a linear scale, instead of a log scale. 
In the bottom panels of Figs.~\ref{fig:MR-fp1} and \ref{fig:Lam-fp1}, we show the relative deviation, $\Delta$, from the fitting formulae. From these figures, we find that the relative deviation from the fitting formulae estimated with QHC21T approaches up to $\sim 20\%$ in a certain stellar model, although the relative deviation expected with the hadronic EOSs is at most $10\%$. This may be a feature of the QHC EOS, which can be distinguished from the hadronic EOSs, even though the $p_1$-mode frequencies are too high to be observed with the current gravitational wave detectors.

In the end, we would like to mention the detectability of the $f$-mode frequencies in gravitational waves from an isolated neutron star. Since our study has been done based on a linear perturbation analysis, we cannot directly estimate the radiation energy of gravitational waves. Nevertheless, one may estimate the corresponding gravitational wave effective amplitude, $h_{\rm eff}$, as 
\begin{equation}
  h_{\rm eff} \sim 3\times 10^{-23}\left(\frac{E_f}{10^{-10}M_\odot}\right)^{1/2}\left(\frac{2\ {\rm kHz}}{f_f}\right)^{1/2}\left(\frac{50\ {\rm kpc}}{D}\right),
\end{equation}
where $E_f$ and $D$ denote the energy in $f$-mode oscillations and the source distance from us \cite{AK1996,AK1998}. Thus, using operating gravitational wave detectors, one may detect the $f$-mode gravitational waves from an isolated neutron star located in our galaxy, if the radiation energy would be more than $10^{-10}M_\odot(\sim 2\times 10^{44}$ erg). We cannot mention typical energy released by gravitational waves from the neutron star oscillations considered here, because it is still theoretically uncertain even though it must strongly depend on astronomical phenomena. For reference, the giant flare from SGR 1806-20 observed in 2004, which is the most energetic ever recorded, is considered to be associated with the neutron star oscillations, where the total (isotropic) released energy in electromagnetic waves is $\sim (3-10)\times 10^{46}$ erg \cite{Israel05}.
For candidates of gravitational wave sources and the estimates of the gravitational wave strain, see, e.g., Ref.~\cite{Andersson21}.

\section{Conclusion}
\label{sec:Conclusion}

How quark matter could appear inside a neutron star is still quite uncertain. In order to see an imprint of the presence of quark matter inside a neutron star, in this study, we examine the $f$- and $p_1$-mode frequencies in gravitational waves from neutron stars by adopting quark-hadron crossover type EOSs, the so-called QHC21T. As a result, we find that the $f$-mode frequencies with QHC21T EOSs are basically smaller than those with Togashi EOS, which is a hadronic EOS and adopted in a lower-density region for constructing QHC21T EOSs, considering neutron stars with the fixed mass. Meanwhile, the $p_1$-mode frequencies with QHC21T EOSs can become larger than those with Togashi EOS in a certain neutron star mass range, considering the neutron star models with the fixed mass. We also find that the universality in the relations for the $f$-mode frequencies multiplied by the stellar mass as a function of the stellar compactness or as a function of the dimensionless tidal deformability can keep even with QHC21T EOSs. 
Thus, this type of universal relation alone does not distinguish QHC21T EOSs from hadronic EOSs.
But, if the mass and $f$-mode frequency are measured, one can use this universal relation to extract the radius and examine how its evolution differs from the typical hadronic trends with gentle stiffening.
We also show the possibility to distinguish QHC21T from hadronic EOSs by using the empirical relations for the $p_1$-mode frequencies multiplied by the stellar mass as a function of the stellar compactness or as a function of the dimensionless tidal deformability. That is, the relative deviation from the empirical relations is at most $10\%$ with hadronic EOSs, but it approaches $\sim 20\%$ with QHC21T in a certain stellar model. 
We note that the $f$-mode gravitational waves from a neutron star located in our galaxy may be detectable with current gravitational wave detectors, if the radiation energy is more than $10^{-10}M_\odot$, while the $p$-mode frequencies must be too high to observe with the current detectors.
In this study, to see the universality in quasinormal modes from neutron stars with QHC EOS, we only adopt the family of QHC21T. Nevertheless, we think the universality in the $f$-mode frequencies shown in Fig.~\ref{fig:MR-Lam-ff} could be held even if one adopts the other type of QHC EOS, because the $f$-mode is a non-nodal global oscillation. On the other hand, how the $p_1$-mode frequencies on the neutron stars constructed with other QHC EOSs deviate from the universal relation derived from the hadronic EOSs, may depend on the type of QHC EOS, because the eigenfunction of the $p_1$-mode oscillations has one node inside the star, i.e., the distribution of the sound speed inside the star must be important for determining the $p_1$-mode frequencies.

\acknowledgments
HS is grateful to H. Togashi for his valuable comments. 
This work is supported in part by Japan Society for the Promotion of Science (JSPS) KAKENHI Grant Numbers 
JP19KK0354  
JP21H01088,  
by FY2023 RIKEN Incentive Research Project,
and by Pioneering Program of RIKEN for Evolution of Matter in the Universe (r-EMU).
T.K. was supported by JSPS KAKENHI Grant No. 23K03377 and by the Graduate Program on Physics for the Universe (GP-PU) at Tohoku University.

\appendix


\begin{thebibliography}{999}
\bibitem{note}  
  We note that QHC19, which is the previous version of QHC21T, is constructed by adopting Togashi EOS for $n_{\rm b}\le 2n_0$ and the quark matter EOS based on the NJL model for $n_{\rm b}\ge 5n_0$.
\end{thebibliography}

\end{document}